\pdfoutput=1
\documentclass{article}
\providecommand{\keywords}[1]{\textbf{\textit{Index terms---}} #1}
\usepackage[utf8]{inputenc}
\usepackage[margin=1in]{geometry}
\usepackage{cite}
\usepackage{amsmath,amssymb,amsfonts}
\usepackage{algorithmic}
\usepackage[ruled,vlined]{algorithm2e}
\usepackage{graphicx}
\usepackage{textcomp}
\begin{document}
\title{Cardiac Adipose Tissue Segmentation via Image-Level Annotations}
\author{Ziyi Huang, Yu Gan, Theresa Lye, Yanchen Liu, Haofeng Zhang,\\ Andrew Laine, Elsa Angelini, and Christine Hendon\footnotemark
}
\footnotetext{This paper is supported in part by NIH-5R01HL14936, NSF-1948540, NIH-4DP2HL127776, New Jersey Health Foundation, and Cheung-Kong Innovation Doctoral Fellowship. (Corresponding author: Christine Hendon, e-mail: cpf2115@columbia.edu)\\
\indent Ziyi Huang, Theresa Lye, Yanchen Liu, and Christine Hendon, are with the Department of Electrical Engineering, Columbia University, New York, NY, 10027, USA\\ 
\indent Yu Gan is with the Department of Biomedical Engineering, Stevens Institute of Technology, Hoboken, NJ, 07030, USA\\ 
\indent Haofeng Zhang is with the Department of Industrial Engineering and Operations Research, Columbia University, New York, NY, 10027, USA\\ 
\indent Andrew Laine is with the Department of Biomedical Engineering, Columbia University, New York, NY, 10027, USA\\ 
\indent Elsa Angelini is with the Department of Biomedical Engineering, Columbia University, New York, NY, 10027, USA; the NIHR Imperial Biomedical Research Centre and ITMAT Data Science Group, Imperial College London, London, UK; and the Telecom Paris, LTCI, Institut Polytechnique de Paris, France} 

\date{}
\maketitle

\begin{abstract}
Automatically identifying the structural substrates underlying cardiac abnormalities can potentially provide real-time guidance for interventional procedures. With the knowledge of cardiac tissue substrates, the treatment of complex arrhythmias such as atrial fibrillation and ventricular tachycardia can be further optimized by detecting arrhythmia substrates to target for treatment (i.e., adipose) and identifying critical structures to avoid. Optical coherence tomography (OCT) is a real-time imaging modality that aids in addressing this need. Existing approaches for cardiac image analysis mainly rely on fully supervised learning techniques, which suffer from the drawback of workload on labor-intensive annotation process of pixel-wise labeling. To lessen the need for pixel-wise labeling, we develop a two-stage deep learning framework for cardiac adipose tissue segmentation using image-level annotations on OCT images of human cardiac substrates. In particular, we integrate class activation mapping with superpixel segmentation to solve the sparse tissue seed challenge raised in cardiac tissue segmentation. Our study bridges the gap between the demand on automatic tissue analysis and the lack of high-quality pixel-wise annotations. To the best of our knowledge, this is the first study that attempts to address cardiac tissue segmentation on OCT images via weakly supervised learning techniques. Within an in-vitro human cardiac OCT dataset, we demonstrate that our weakly supervised approach on image-level annotations achieves comparable performance as fully supervised methods trained on pixel-wise annotations.

\end{abstract}


\section{Introduction}
\label{sec:introduction}
Cardiovascular disease is the leading cause of death in the United States, with atrial fibrillation alone affecting at least 2.3 million people\cite{khurshid2018frequency}. Treatment of complex arrhythmias such as atrial fibrillation and ventricular tachycardia is through catheter ablation, which directly destroys the cardiac substrates that cause irregular impulse propagation. However, this treatment is sub-optimal, due to the lack of capability to accurately identify optimal ablation targets. With the knowledge of patients’ heart structure, the ablation strategy can be further optimized by avoiding critical structures and identifying arrhythmia substrates, such as areas with increased amounts of adipose tissues. Recent work has shown that an increased amount of adipose tissues within the myocardium is a substrate for cardiac arrhythmias \cite{samanta2016role,bonou2021cardiac,pabon2018linking,el2021posterior,sung2020personalized}.

Optical coherence tomography (OCT) is a non-destructive optical imaging modality that has the capability to capture myocardial structures such as Purkinje network \cite{yao2016myocardial}, atrial ventricular nodes\cite{gupta2002imaging}, sinoatrial nodes\cite{ambrosi2012quantification}, and myofiber organization\cite{goergen2012optical}. In addition, it can be used to resolve critical tissue substrates of arrhythmias, such as fibrosis and adipose tissues \cite{lye2019imaging}. With the development of OCT-integrated catheters\cite{fleming2010real}, OCT can image the heart wall in real time through percutaneous access \cite{wang2011vivo}, which holds promise to
aid catheter ablation.

To benefit from the real-time capacity of OCT imaging,
analysis of OCT images is expected to be automated for timely decision making. Evaluation of adipose tissue distribution within a human atrial sample requires pixel-wise analysis of large volumetric datasets \cite{gan2019characterization}. Manually annotating adipose tissues within a single OCT volumes can take a well-trained annotator over 10 hours. Therefore, automated identification of cardiac tissues, especially adipose tissue, in OCT images is greatly needed.

Current automated analysis on cardiac OCT images is mostly based on fully supervised learning models\cite{9600862,huang2020heterogeneity,huang2020segmentation}. These models were limited and suffered from the drawback of manual workload in the labeling process. To avoid overfitting, a large amount of data is required to support the model training. For segmentation tasks, the labeling process is extremely time-consuming and has limited accuracy. Moreover, OCT images are volumetric, adding an additional challenge to labeling. Thus, automatic analysis with weakly supervised learning models is of great interest.

\begin{figure*}
\begin{center}
\includegraphics[width=0.9\textwidth]{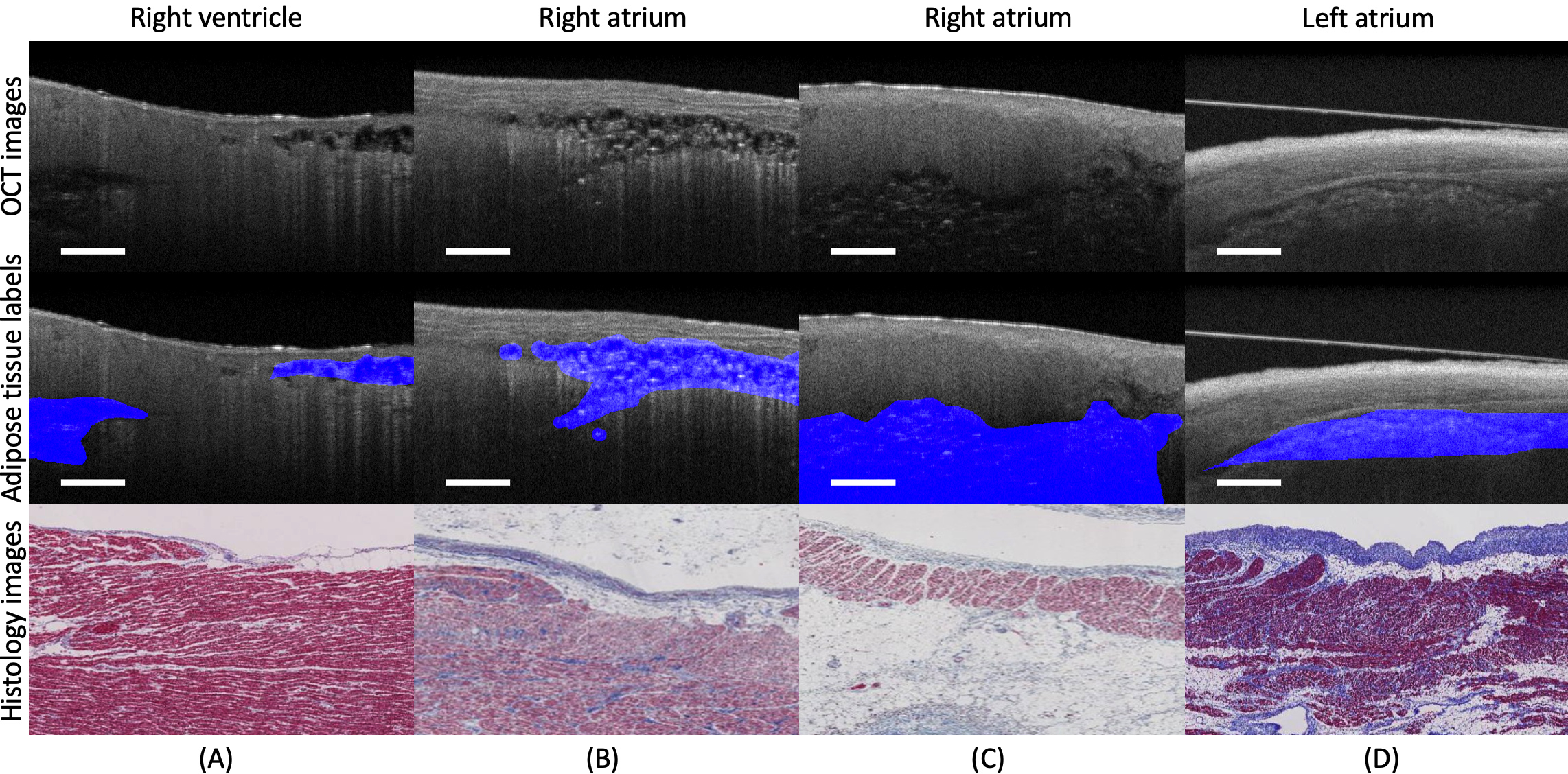}
\end{center}
\vspace{-1em}
\caption{Representative OCT images from cardiac dataset. Sample (A) is obtained from right ventricle. Sample (B) and sample (C) are obtained from right atrium. Sample (D) is obtained from left atrium submerged in PBS solution. The features of adipose tissue present great variations among different locations and imaging conditions. The unclear boundary and irregular shape of adipose tissues add unique challenges for automated segmentation. Scale bar: $500 \  \mu m$. } 
\vspace{-1em}
\label{challenge}
\end{figure*}

Although recent study has investigated in retinal OCT analysis \cite{wang2021weakly,yoo2021feasibility}, transferring OCT retinal segmentation to cardiac  solution for weakly supervised cardiac OCT segmentation is still elusive for three reasons. First, cardiac adipose and fibrosis tissues can appear in multiple sub-regions with irregular shapes and infiltrating patterns. Thus, the cardiac OCT images are more complicate than retinal substrates with rather regular layered structure. Second, boundaries between cardiac substrates are more blurry than between the retinal layers. Third, cardiac substrates have larger variance among patients than retinal tissues.

In this study, we present a weakly supervised learning framework for cardiac tissue segmentation using image-level labels. Our training approach has two stages, namely pseudo label generation and segmentation network training. We first use the class activation map (CAM) results obtained from a binary classification network to generate adipose location seeds. Then, we develop a superpixel-based segmentation algorithm to generate pseudo labels followed by segmentation training. Our contributions are as follows:\\
(1) We propose a weakly supervised learning framework for cardiac tissue segmentation. Our model is trained without the need of pre-training or domain adaptive learning. \\
(2) We combine CAM with superpixel segmentation to effectively address the sparse seed challenge caused by irregular shape and unclear boundary of adipose tissues.\\ 
(3) We evaluate our approach on a human cardiac dataset and demonstrate that our weakly supervised model achieves comparable performance with fully supervised algorithms.

\begin{figure*}[t]
\begin{center}
\includegraphics[width=0.9\textwidth]{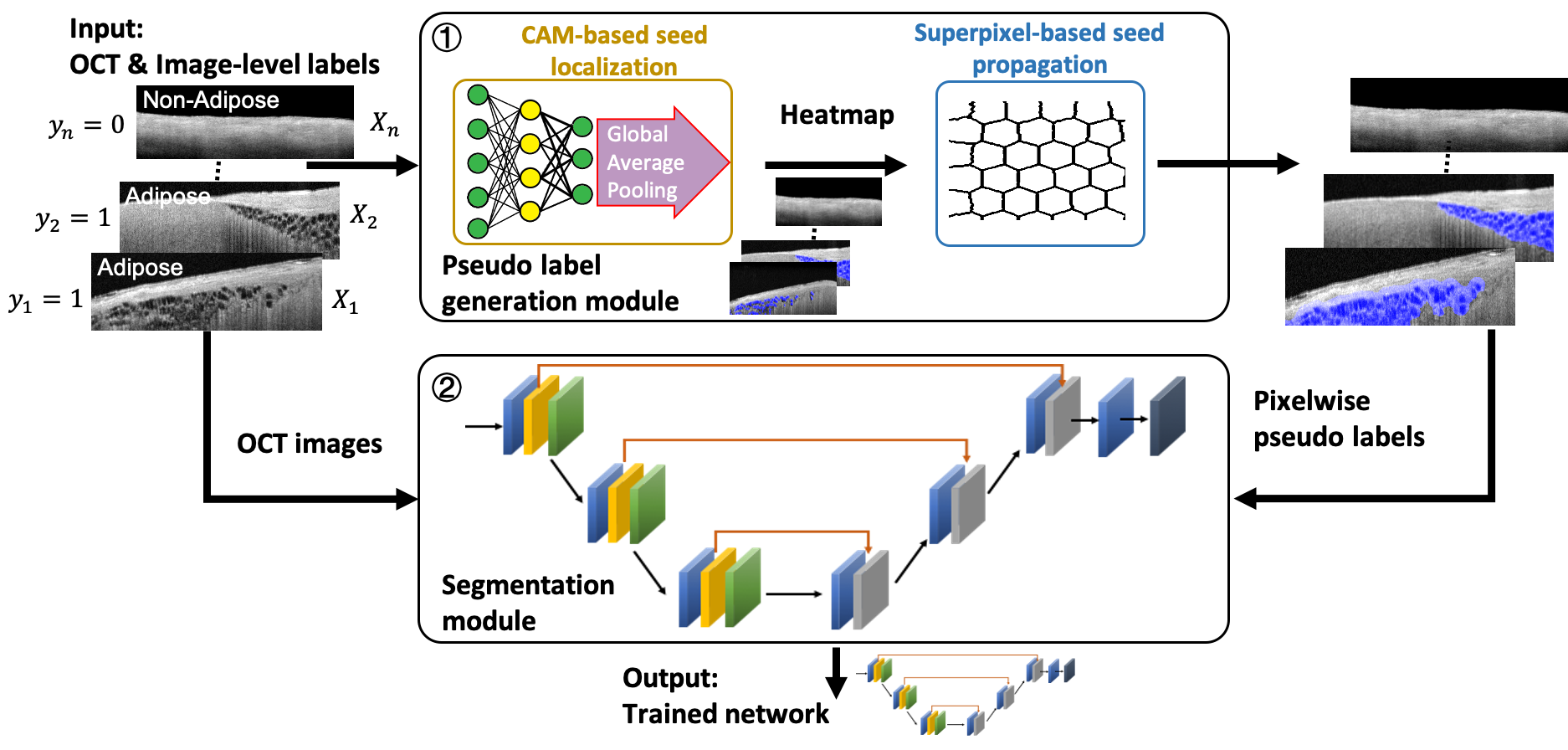}
\end{center}
\vspace{-1em}
\caption{Algorithm training flow of the proposed weakly supervised segmentation approach. The framework consists of two separate modules, namely pseudo label generation and segmentation network training. In pseudo label generation module, pixel-wise pseudo annotations are generated by the integration of CAM and superpixel methods. In segmentation network training module, a segmentation network is trained on the pseudo labels with a novel loss function.} 
\vspace{-1em}
\label{flow}
\end{figure*}

\section{Related Work}
Regarding tissue analysis on cardiac OCT images, \cite{gan2016automated} imaged and analyzed features on dense collagen, loose collagen, fibrotic myocardium, normal myocardium, and adipose tissue for automatic classification. In \cite{meiniel2016sparsity}, segmentation was obtained from the variance map through compressive sensing reconstruction. In \cite{lye2019imaging}, the distributions of adipose tissues and fiber orientations were retracted and mapped throughout human left atrium, while in \cite{ambrosi2009virtual}, the visualization of cardiac fibers in the atrium, ventricle, atrioventricular node, and sinoatrial node were presented. Overall, conventional cardiac OCT image analysis relies on handcrafted features for tissue characterization or fiber orientation-based methods to focus on myofibers. 

Deep learning approaches have achieved great success in OCT image segmentation tasks \cite{islam2020deep,mirshahi2021foveal,guo2019automatic,masood2019automatic,pekala2019deep,lo2020microvasculature,lee2021automated}. \cite{PEKALA2019103445} developed a fully convolutional network with Gaussian process based post processing  for retinal OCT segmentation. \cite{fang2017automatic} proposed a novel framework which combined a hybrid convolutional neural network and graph search method for retinal layer boundary detection. \cite{alam2020av} developed a fully convolutional network based AV-Net for artery-vein classification. Their model contained a multi-modal training process which involved both en-face OCT and optical coherence tomography angiography (OCTA) to provide the intensity and geometric profiles. \cite{huang2020heterogeneity} trained a fully supervised segmentation network for cardiac tissue segmentation and used model uncertainty to estimate tissue heterogeneity. Existing work mainly relies upon the fully supervised learning. 

In contrast to fully supervised methods, weakly supervised approaches use higher level labels to guide the pixel-level segmentation training process. \cite{wang2021weakly} successfully segmented lesions by calculating the differences between the input abnormal images and normal-like retinal OCT images from a CycleGAN model. \cite{yoo2021feasibility} employed a few shot learning technique for retinal disease classification and applied a GAN to enrich normal OCT images with OCT images of rare diseases. \cite{jiang2020weakly} proposed a Noise2Noise \cite{lehtinen2018noise2noise} based weakly supervised learning model for OCTA image reconstruction task. 

\section{Problem Analysis}
 Our study is conducted on a cardiac dataset that was acquired from 44 human hearts with median age at 62 years. The dataset contains both healthy hearts, end-stage heart failure, atrial fibrillation, coronary heart disease, cardiomyopathy, and myocardial infarction. A detailed clinical characteristic is presented in Section \ref{dataset}. These various disease conditions might alter the visual features of cardiac substrates, raising the following unique challenges on the algorithm design:

\textbf{Various features and irregular shapes.} As shown in Fig. \ref{challenge}, adipose regions present great variations among cardiac OCT images from human donors with cardiovascular disease due to heterogeneous heart remodeling. In Fig. \ref{challenge} (A), intra-scan inconsistency can be clearly observed in the two sub-regions. Meanwhile, in comparison with Fig. \ref{challenge} (B), the size of fat cells in Fig. \ref{challenge} (A) is much smaller and the number of fat cells is larger, as indicated in the histology images. In addition, the distance to the endocardium tissue can also affect tissue appearance. Adipose tissues in Fig. \ref{challenge} (C) are deeper in the myocardium and appear darker and blurrier than in Fig. \ref{challenge} (B). Finally, the features and shapes can be further impacted by experimental conditions. In Fig. \ref{challenge} (D), the OCT image was obtained from tissues submerged in phosphate buffered saline (PBS). In this sample, the adipose tissues have very low contrast with the surrounding normal tissues. 


\textbf{Similar pattern among adipose tissue and noise.} Image noise and artifacts are inevitable during the acquisition process. Features of adipose tissue are very similar to those of speckle noise and artifacts.

\textbf{Data imbalance and limited training data.} In human cardiac samples, the majority of regions are normal tissues, such as myocardium and endocardium, rather than targeted adipose tissues. In our dataset, only $10\%$ OCT images shows large clusters of adipose tissues. At pixel level, pixels belonging to adipose tissues only account for 2.6\% of the total pixels to label. Even for images that contain adipose tissues, the ratio of number of adipose-related pixels over total number of pixels is very small. Hence, the samples that are informative for model training are very limited.

\section{Methodology}\label{sec:meth}

In this paper, we propose a weakly supervised learning framework for cardiac tissue segmentation task. We denote the space of images by $\mathcal{X}$. For any image $X_i \in \mathcal{X},\  i = 1, ..., n$, the image-level annotation $y_i \in \{0,1\}$, indicates whether $X_i$ contains the adipose tissues. Figure \ref{flow} shows the pipeline of our proposed framework. As shown, our training approach consists of two major stages: pseudo label generation and segmentation network training. The pseudo label generation module is formed by two components: we first apply the CAM approach to generate initial adipose seeds and then we use superpixel-base segmentation method to propagate the adipose seeds into pseudo pixel-wise labels. A detailed pseudo algorithm for the pseudo label generation module is listed in Algorithm \ref{algo1}. In the segmentation module, we introduce a novel loss function with a special focus on the adipose seed regions to increase the detection performance of our segmentation network.

\subsection{Pseudo Label Generation}
\subsubsection{CAM-based seed localization} Image-level labels do not include any location clue for the target tissue. Thus, it cannot be directly used to train the segmentation networks. So, the first stage of our model is to find reliable adipose seeds to indicate the location of adipose tissues. We use the class activation map to generate initial adipose seeds. Since the adipose tissues do not have regular shapes and may appear at multiple cluster regions, we employ the global average pooling (GAP) layer to generate the class activation maps, as it has advantages on identifying the extent of target tissue regions over the global max pooling layer \cite{zhou2016learning}. 

The class activation maps have strong responses on regions with artifacts and high intensity noise. To increase the reliability of pseudo label generation, we apply a boundary masking algorithm on the class activation maps to filter out adipose seeds that are located in the background regions (false positive caused by noise) and regions close to the tissue-background boundary (false positive caused by artifacts). We adapt the cardiac layer segmentation algorithm from \cite{gan2016automated} and use the boundary of the top generated layer as the tissue-background interface. After getting the tissue surface, we remove adipose seeds above or close to the tissue-background boundary. 

\subsubsection{Superpixel-based seed propagation} 

Superpixels are generated as in \cite{achanta2010slic}. An entire superpixel is labeled as adipose tissue if one of its inner pixels is labeled as adipose tissue. Upon the generation of superpixels, the initial segmentation pseudo labels can be further improved to eliminate the following two misclassifications: 1) The adipose seeds may omit some adipose regions and 2) the adipose seeds may incorrectly mark the normal regions as the adipose regions, due to the artifacts and intensity noise. To further remove the noisy annotations, we apply the Markov spatial regularisation strategy to add the ignored regions and remove the noisy adipose superpixels which only contain the normal tissues. Since the adipose cells are clustered in the cardiac tissue, the neighbors of an adipose region are more likely to belong to the adipose tissue clan, while small isolated adipose superpixels are more likely to be the normal regions corrupted by noise. Based on these criteria, we develop a simple yet effective spatial regularisation strategy: the label of a superpixel will be updated if most of its neighbors ($\geq 80\%$) belong to another class.

\begin{algorithm}[t] \label{algo1}
\SetAlgoLined
\textbf{Input}: Training dataset $\mathcal{X}=\{X_1 , X_2 , . . . , X_n \}$ with image-level labels $\mathcal{Y}=\{y_1 , y_2 , . . . , y_n\}$;\\
\textbf{Output}: Pixel-wises pseudo labels.\\
\textbf{Procedure}:\\ 
Step 1: Train localization network from $\mathcal{X}$ and $\mathcal{Y}$ .\\
Step 2: Apply CAM method to generate the initial tissue seed results $\mathcal{C}=\{c_1 , c_2 , . . . , c_n\}$.\\
Step 3: Apply the boundary masking on $\mathcal{C}$ and get updated tissue seeds $\hat{ \mathcal{C} }$.\\
Step 4: Apply superpixel-based propagation method on $\hat{ \mathcal{C} }$ to generate the initial pseudo segmentation labels $\mathcal{S}$.\\
Step 5: Update $\mathcal{S}$ with the spatial regularisation strategy and get the final pseudo segmentation labels $\hat{\mathcal{S}}$.

\caption{Algorithm Framework for Pseudo Label Generation Module}

\end{algorithm}

\subsection{Segmentation Network Training}
Adipose tissues are sparse, in comparison with normal tissues, such as myocardium and endocardium. Without special consideration, the segmentation  performance might be severely limited due to data imbalance. A traditional way to solve data imbalance issue is to add special weights on the minority classes. However, in our initial training data, the pseudo segmentation maps generated by CAM-superpixels are not very precise, and thus cannot support the class-weighting strategy. To overcome this challenge, we use a seed loss, inspired by \cite{kolesnikov2016seed}, to optimize our segmentation network.

First, we denote $p_k(x)$ as the predicted probability for class $k$ at the pixel position $\textit{x}\in\Omega$ with $\Omega\subset \mathbb{R}^2$ and $y_k(x)$ as the one-hot encoding of the ground-truth annotation for class $k$, where $k\in \{1,2,...,K\}$ and $K$ is the number of classes. In this work, $K=2$.
The cross entropy loss (CEL) and seed loss (SL) is defined as follows:

\begin{equation}\label{loss_ce}
      CEL = -\frac{1}{|\Omega|} \sum_{x\in \Omega} \sum_{k=1}^K y_k(x) \log(p_k(x)) 
\end{equation}

\begin{equation}\label{loss_fl}
      SL = -\frac{1}{|\Omega_1|} \sum_{x\in \Omega_1} y_1(x) \log(p_1(x)) = -\frac{1}{|\Omega_1|} \sum_{x\in \Omega_1} \log(p_1(x))
\end{equation}
where $\Omega_1=\{x\in \Omega: y_1(x)=1\}$ is the set of locations that are labeled with class $1$ (i.e., the adipose class).
Compared with the CEL (Eq.\ref{loss_ce}), the SL (Eq.\ref{loss_fl}) only focuses on the regions of adipose tissue, and thus, it helps to reduce the impact of the false negatives in the pseudo segmentation labels. 

We also use Dice loss (DL) in our loss function to learn the context information. The DL is defined as:
\begin{equation}
\begin{array}{cccccccccc}
    DL &=& 1- \frac{1}{K}\sum\limits_{k=1}^K\frac{2\sum\limits_{x \in \Omega}(p_k(x)y_k(x))}{\sum\limits_{x \in \Omega}(p_k(x))^2  + \sum\limits_{x \in \Omega}(y_k(x))^2}\\
\end{array}
\end{equation}
 Finally, our segmentation network is jointly optimized by the combination of CEL, SL, and DL:
\begin{equation}\label{loss_func}
      Loss = w_1 \cdot CEL + (1 - w_1) \cdot SL + w_2 DL
\end{equation}
where $w_1$ and $w_2$ are weight hyper-parameters. 

\begin{table}[t]
\centering
\caption{Clinical characteristics of heart donors}
\begin{tabular}{lcccc}
\hline
Characteristic & Value   \\ \hline
N  & 44 \\ 
Demographic profile &\\
\ \ Age in years, median (average)  & 62 (62.2) \\
\ \ Female, n (\%) & 20 (45.5)\\ \hline
Medical history, n (\%) & \\
\ \ Heart failure & 10 (22.7) \\ 
\ \ Cardiomyopathy & 8 (18.2)\\
\ \ Coronary artery disease & 11 (25.0) \\
\ \ Myocardial infarction & 10 (22.7) \\
\ \ Atrial fibrillation & 3 (6.8)\\ 
\ \ Chronic obstructive pulmonary disease &16 (36.4)\\
\ \ Diabetes & 17 (38.6)\\
\ \ Hypertension & 27 (61.4) \\ \hline
Cause of death, n (\%)\\
\ \ Cardiac arrest & 18 (40.9) \\
\ \ Cardiopulmonary arrest & 2 (4.5)\\
\ \ Respiratory failure  & 5 (11.4)\\
\ \ Chronic obstructive pulmonary disease &  1 (2.27)\\
\ \ Congestive heart failure & 1 (2.27) \\
\ \ Others, cardiac related & 11 (25.0) \\
\ \ Others, not cardiac related & 6 (13.6)\\ \hline

\end{tabular}
\label{tab_honourable}
\end{table}





\begin{figure}[t]
\begin{center}
\includegraphics[width=0.9\textwidth]{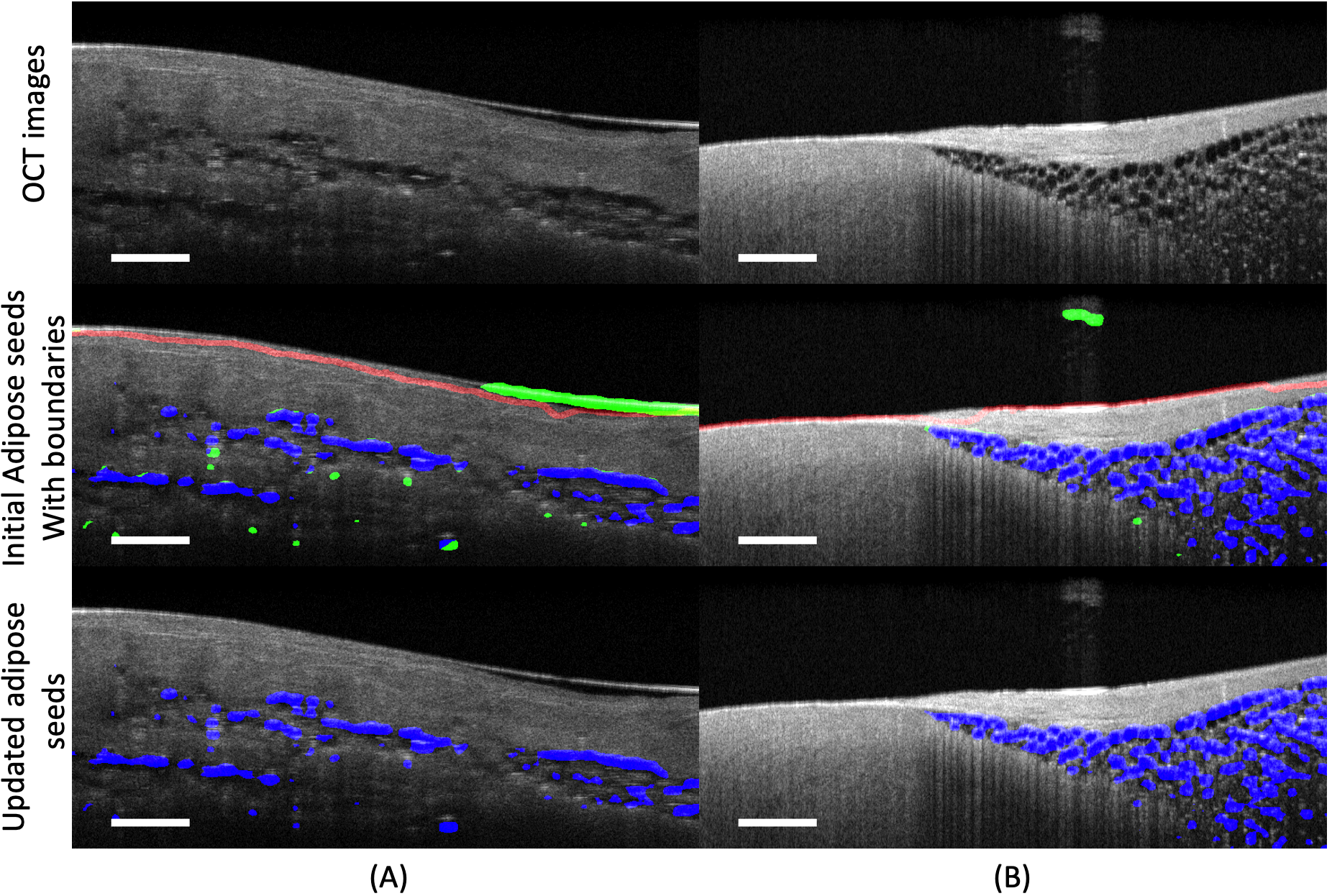}
\end{center}

\caption{Comparison of tissue seeds before and after the boundary masking algorithm. Red: the detected tissue-background boundaries; blue: accurately annotated adipose seeds; green: false positives. As shown, the boundary masking algorithm can effectively remove the false positive adipose seeds caused by the artifacts and noise. Benefiting from it, the adipose seeds are more precise to be propagated for segmentation guidance. Scale bar: $500 \  \mu m$.} 
\label{cam2}
\end{figure}

\begin{figure}[t]
\begin{center}
\includegraphics[width=0.9\textwidth]{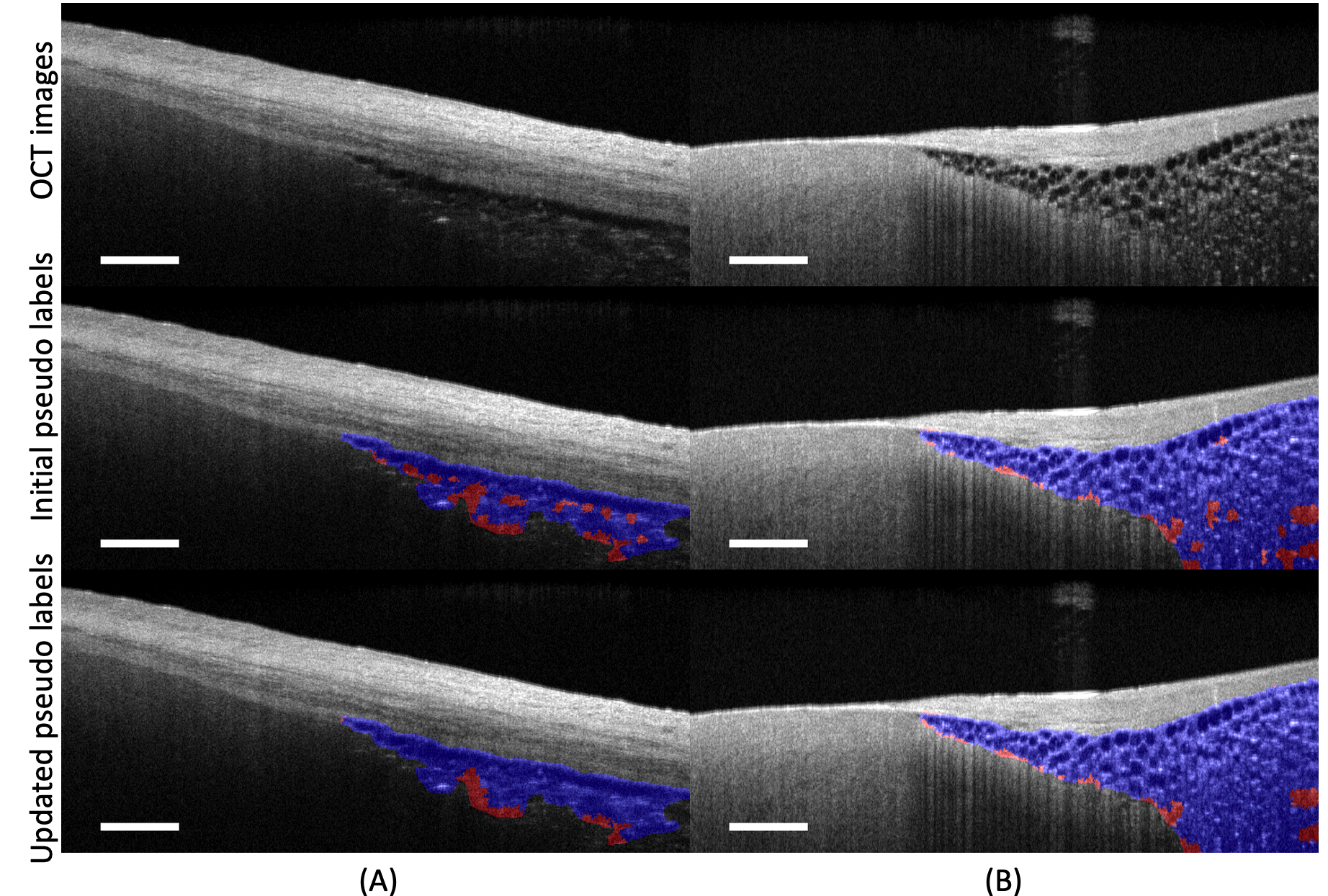}
\end{center}

\caption{Comparison of pseudo labels with and without the spatial regularisation strategy. Blue: accurately annotated adipose pixels; red: false negatives. The spatial regularisation strategy helps to correct the mis-labeled pseudo labels by using the context information from nearby regions. After applying it, the false negatives have been significantly reduced. Scale bar: $500 \  \mu m$.} 
\label{fig2}
\end{figure}

\begin{table}[t]

\centering
\caption{Evaluation metrics (\%) of adipose tissue seeds before and after the boundary masking algorithm. }
\begin{tabular}{cccc}
\hline
& & Before & After   \\  \hline 
{Accuracy} & &80.79 $\pm$ 1.15 & 83.91 $\pm$ 2.45  \\ \hline
{Precision} & &56.43 $\pm$ 11.95 & 75.90 $\pm$ 8.18  \\ \hline

\end{tabular}

\label{tb_seed}
\end{table}
\begin{table*}[t]
\centering
\caption{Evaluation metrics (\%) on tissue pseudo labels before and after the Markov spatial regularisation.}
\begin{tabular}{cccc}
\hline
Method & True Positive Rate & False Positive Rate & Dice Coefficient   \\ \hline 
Superpixel  & 71.17 $\pm$ 6.36 & 10.03 $\pm$ 2.87& 67.09 $\pm$ 2.96 \\ \hline
Superpixel + Spatial regularisation  & 71.77 $\pm$ 6.72 & 8.33 $\pm$ 2.90 & 69.70 $\pm$ 3.34\\ \hline

\end{tabular}
\label{tab3}
\end{table*}

\begin{table*}[h]
\centering
\caption{Evaluation metrics (\%) of different models on whole dataset. }
\begin{tabular}{cccc}
\hline
Method & True Positive Rate & False Positive Rate & Dice Coefficient   \\ \hline 
 U-Net (Fully Supervised)  & 83.73 $\pm$ 3.930 & 3.62 $\pm$ 1.05 & 80.53 $\pm$ 8.09  \\ \hline
 Proposed & 86.05 $\pm$ 5.534 & 6.73 $\pm$ 2.63 & 79.67 $\pm$ 6.98  \\ \hline
w/o boundary masking & 72.63 $\pm$ 14.97 & 4.55 $\pm$ 2.07 & 73.56 $\pm$ 11.31  \\ \hline

w/o Markov spatial regularisation & 78.02 $\pm$ 7.90 & 4.47 $\pm$ 1.62 & 77.24 $\pm$ 6.31  \\ \hline
Seed loss + Dice loss & 88.38 $\pm$ 8.79 & 9.41 $\pm$ 2.07 & 73.35 $\pm$ 9.24  \\ \hline
CE loss + Dice loss& 74.63 $\pm$ 15.77 & 3.47 $\pm$ 0.71 & 75.94 $\pm$ 10.74 \\ \hline

\end{tabular}
\label{tab1}
\end{table*}

\begin{figure*}[h]
\begin{center}
\includegraphics[width=0.9\textwidth]{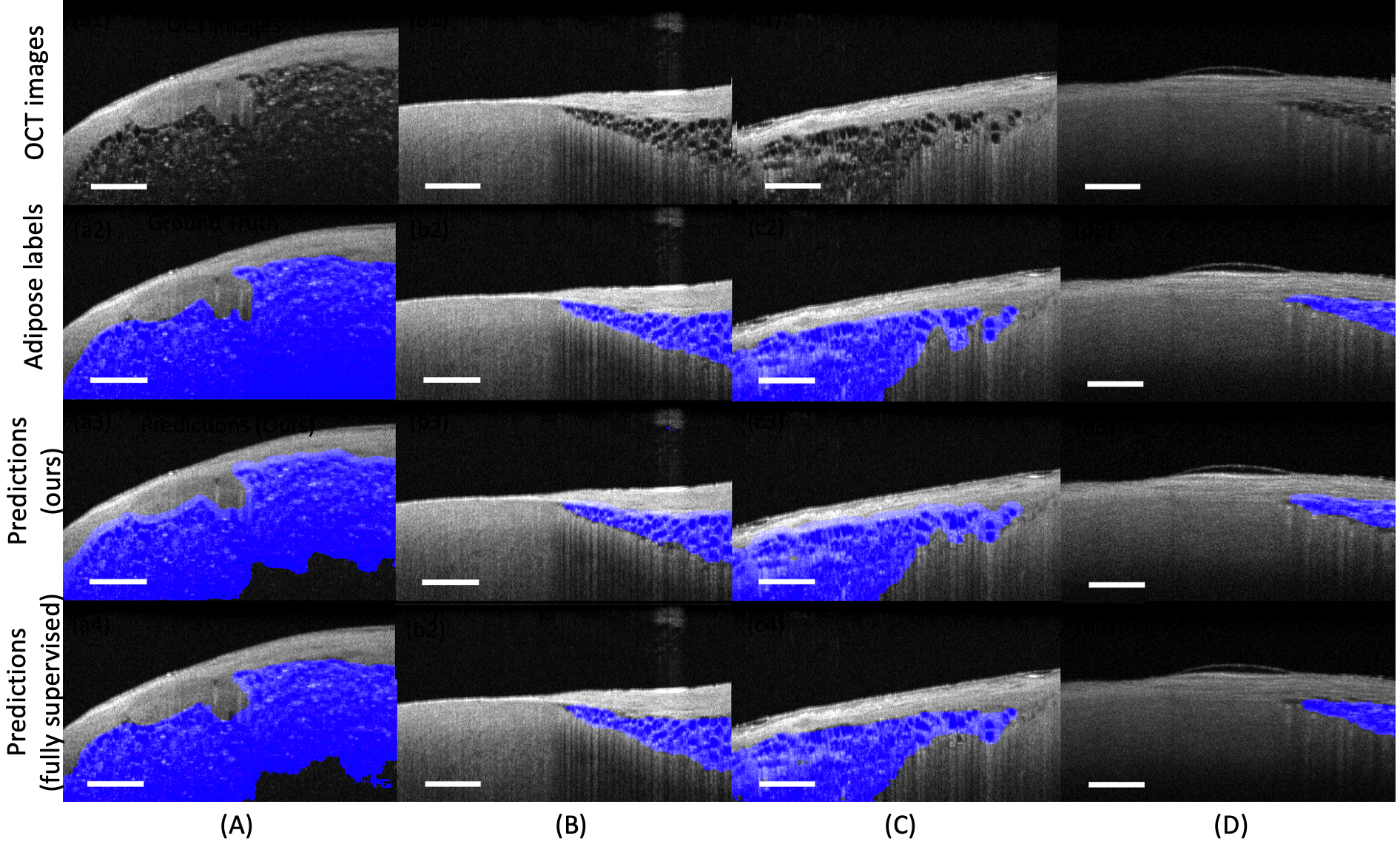}
\end{center}
\vspace{-1em}
\caption{Representative segmentation results from human atrium and ventricle samples. Our proposed approach accurately identifies the adipose tissues located at different regions with various sizes and shapes. All prediction results are highly consistent with the ground truth labels. Scale bar: $500 \  \mu m$} 
\vspace{-1em}
\label{seg}
\end{figure*}

\begin{figure}[h]
\begin{center}
\includegraphics[width=0.9\textwidth]{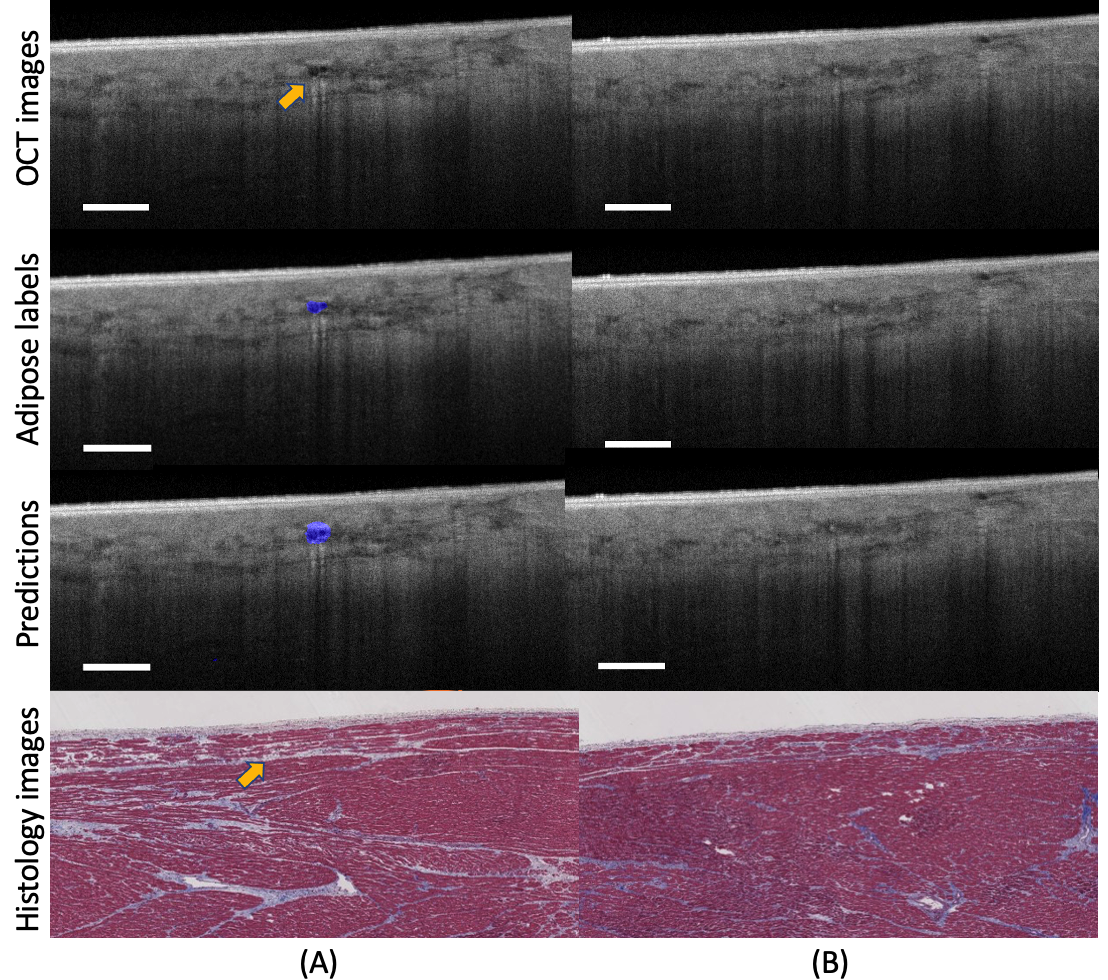}
\end{center}

\caption{The prediction results of images obtained from nearby regions. Our approach successfully pinpoints the adipose tissues from other tissue types, showing its strong identification ability on adipose tissues. Scale bar: 500 $\mu m$.} 
\vspace{-1em}
\label{fib}
\end{figure}
\begin{figure*}[h]
\begin{center}
\includegraphics[width=0.9\textwidth]{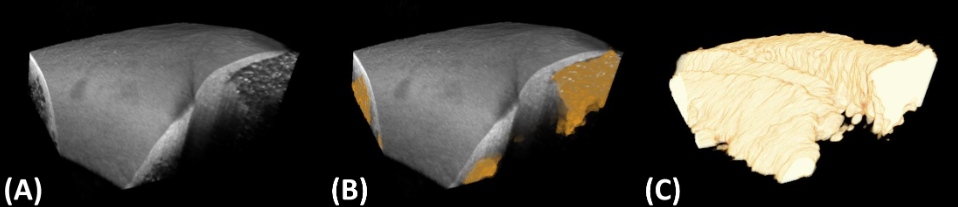}
\end{center}

\caption{3D visualization of adipose tissue segmentation. (A): the original OCT volume; (B): the original volume overlaid with segmented adipose regions; (C): the segmented adipose regions from proposed approach. The segmented boundaries accurately delineate the morphological changes in adipose shape. }
\label{fib:3D}
\end{figure*}

\section{Experiment Evaluation}

\subsection{Dataset}\label{dataset}
We evaluate the performance of our proposed model on the human cardiac dataset previously used in \cite{gan2019characterization}. It consists of \textit{in-vitro} cohort of 385 images taken from 44 human atria and ventricles using the Thorlabs OCT system. The samples were acquired through a National Disease Research Interchange approved protocol from Columbia University. All specimens were de-identified and considered not human subjects research, according to the Columbia University Institutional Review Board under 45 CFR 46. Table \ref{tab_honourable} presents the detailed clinical characteristic of the human donor hearts. Each OCT image is of size 512 $\times$ 800 pixels with a field of view of 2.51 mm $\times$ 4 mm. Three experts, blinded to the algorithm design, annotated the OCT images under the guidance from a pathologist. All images were carefully annotated at pixel level with visual cron-check on corresponding histology images. Our evaluation is conducted on a five-fold cross validation strategy with validation sets randomly divided over human subjects. 

\subsection{Implementation Details}
\textbf{Seed localization network.} To avoid overfitting, we only train a localization network with three hidden layers for adipose tissue seed generation. We use the ReLU function as the activation function. The number of neurons in each hidden layer is 32, 32, and 64. We use GAP layer as the final layer on the localization network to learn the cluster pattern of adipose tissues. The networks are optimized on cross entropy loss via Adam optimizer \cite{kingma2014adam} with random Glorot uniform initialization \cite{glorot2010understanding}. Over the cross validation sets, all networks converged within 300 epochs with a learning rate of $1e^{-3}$. 

\textbf{Segmentation network.} We employ the classic medical segmentation network UNet \cite{ronneberger2015u} as the baseline of our learning framework. The hyper-parameter $\omega_1$ in the loss function (Eq.\ref{loss_func}) was determined from $[0,1]$ according to the proportion of adipose tissues in the training set with $\omega_2 = 0.5$. All segmentation networks were randomly initialized and converged within 200 epochs with a learning rate of $1e^{-3}$ and weight decay $10^{-4}$.

\subsection{Evaluation Metrics}
In our experiment, we use accuracy and precision to evaluate the overall accuracy and the detection performance of adipose seed results. For pseudo label generation and segmentation evaluation, we use true positive rate (detection rate), false positive rate, and Dice coefficient (F1 score) to evaluate the tissue segmentation performance. 





\subsection{Evaluation of Pseudo Label Generation}

\textbf{Adipose tissue seed localization.} The binary accuracy for our proposed localization network achieves very stable performance ($ >= 95\%$) on all validation sets. Figure \ref{cam2} presents two representative adipose seed results generated from our localization network. In Fig. \ref{cam2}, the detected tissue-background boundary is delineated in red and the accurately located adipose seeds are marked in blue with mis-classified adipose seeds marked in green. As seen, the boundary masking algorithm can effectively remove the misclassified edges and background noise from the original adipose seed results and the adipose tissues are successfully marked with same seeds. In Table \ref{tb_seed}, we report the accuracy and precision of adipose seeds before and after boundary masking. As shown, both accuracy and precision are improved after applying the boundary masking algorithm. In particular, the precision of adipose seeds has been significantly increased by approximately 20\%, which evidently demonstrates the effectiveness of boundary masking on accuracy improvement.

\textbf{Pseudo label generation.} 
Figure \ref{fig2} shows two superpixel segmentation results with accurately segmented pixels marked in blue and false negatives marked in red. In Table \ref{tab3}, we provide a quantitative evaluation of our spatial regularisation strategy. There is an increase of $2\%$ in Dice coefficient after applying it.

\subsection{Comparison with Fully Supervised Segmentation}
\subsubsection{Cross Validation Experiments}
We use fully supervised models trained from pixel-wise accurate segmentation masks as the baseline for comparison. Table \ref{tab1} summarizes the averaged results and the standard deviation of our proposed weakly supervised approach and fully supervised baselines. 

\textbf{Weakly supervised learning vs fully supervised learning.} Our weakly supervised model, trained from image-level labels, achieves comparable quality results than the fully supervised model that trained on pixel-wise labels. In addition, our dataset was acquired within a time frame that spanned over five years. During this time frame, imaging setup, such as sample freshness, imaging condition, and tissue preparation, varied among experiments. Thus, our results also demonstrate the generalization ability of our model against imaging condition variance, showing its strong potential on real-world clinical applications.

\textbf{Ablation study.} Performance drops are observed in Table \ref{tab1} when the boundary masking algorithm and spatial regularisation strategy are removed. In particular, after removing the boundary masking algorithm, the true positive rate is severely decreased along with an increased standard deviation. This result indicates the necessity of using boundary masking algorithm to improve adipose seed quality at the early stage of pseudo label generation. In comparison with the false positive rate, the true positive rate has notable changes after applying the spatial regularisation, which shows its effectiveness on false negative correction. 

We further conduct experiments to assess the influence of different loss functions in our proposed model. As shown in Table \ref{tab1}, the use of seed loss can notably increase the model detection performance but meanwhile hinder the false positive rate. In contrast, the cross entropy loss is more efficient on controlling the false alrams. These results show that the use of seed loss can efficiently reduce the impact of false negatives in the pseudo labels. This detection rate and false alarm trade-off can be balanced through adjusting the weights of seed loss and cross entropy loss.

\subsubsection{Representative Segmentation Results}
In this section, we present the visual output of our proposed weakly supervised model in overall performance, small adipose tissue region detection, and 3D segmentation.

\textbf{Overall performance.} Figure \ref{seg} shows the predicted tissue maps on four human cardiac samples. In Fig. \ref{seg} (A) and (B), our model accurately localizes the adipose tissue regions in arbitrary shapes. Meanwhile, in Fig. \ref{seg} (A), we can also observe the over segmented regions (regions at left corner) in the ground truth figure. Human annotators tend to oversegment the regions below the penetration depth, while for the network, it may identify these regions as non-adipose tissues because of the low signal to noise ratio. This over-segmentation tendency can lead to the decreased values in evaluation metrics. In Fig. \ref{seg} (C), our model successfully identifies the adipose tissues in the multiple regions with different penetration depth. In Fig. \ref{seg} (D), we show a human atrium sample which is slightly off the focus. Similar to previous results, our models still accurately differentiates the adipose tissues from other tissues, showing its robustness over different image qualities. In all cases, the predicted results are highly consistent with the ground truth labels. These results demonstrate the learning ability of our model via image-level labels, showing its effectiveness on clinical tissue identification.

\textbf{Identifying small adipose tissue regions.} Figure \ref{fib} presents two images obtained from nearby regions within the same human heart. As shown, Fig. \ref{fib} (A) and (B) are very similar and they all contain large regions of fibrosis tissue. However, in Fig. \ref{fib} (A), there is a small cluster of adipose tissue surrounded by the fibrosis tissues, while in Fig. \ref{fib} (B), there is no adipose tissue. As shown, this is a very challenging segmentation task due to the size of adipose tissue and the blurry boundary between different tissue types. Our model accurately delineates the adipose tissue regions in Fig. \ref{fib} (A) and it does not put any false alarm in Fig. \ref{fib} (B). In both cases, our model successfully distinguishes adipose tissues from other cardiac substrates. These results further demonstrate the strong learning ability of our model, as it can learn the most discriminative features via image-level labels, rather than simply memorizing the training samples.

\textbf{Visualization of 3D segmentation.} Figure \ref{fib:3D} shows a typical result of 3D visualization of adipose tissue segmentation. We sequentially apply the trained network to segment consecutive Bscans and align segmented Bscans in 3D space. As shown, the segmented boundaries accurately delineate the morphological changes in the adipose tissues. Even though our model is trained on a small quantity of training data with image-level labels, it still successfully segments adipose regions of various sizes. These results indicate that our model has great potential to be applied to assess adipose tissue regions in catheter-based ablation operations.

\section{Discussion}
In this study, we propose a weakly supervised learning framework for cardiac adipose tissue segmentation on OCT images. Our approach contains two powerful modules: the pseudo label generation and the segmentation network training. In the pseudo label generation module, we use the superpixel-based propagation algorithm to address the sparse location seed challenge raised in the CAM results. Benefiting from our boundary masking algorithm and spatial regularisation strategy, the quality of the pseudo labels has been significantly improved for the training guidance. In the segmentation network training module, we introduce a novel loss function to increase the adipose tissue detection performance. By evaluating on a human cardiac dataset with cross validation strategy, our model achieves comparable results with the fully supervised baseline, showing its effectiveness on tissue characterization. Our study bridges the gap between the demand on automatic tissue analysis and the lack of high-quality pixel-wise annotations. To the best of our knowledge, this is the first study that attempts to address cardiac tissue segmentation via weakly supervised learning techniques. 

One limitation of this study is that the results are evaluated on a benchtop OCT system. To move towards the aid of ablation procedures, catheter-based OCT system is needed, as it can help to optimize the treatment strategy by providing real-time cardiac substrates information. In the future, we will extend our current work into catheter-based \textit{in-vivo} OCT images. Such extension will require further investigation on challenges such as image quality degeneration and motion disturbance. Compared with the benchtop OCT images, catheter-based OCT images are with lower image quality, suffering from lower contrast and motion effects. Without special consideration, the decreased image contrast may hinder the performance of the model. Motion disturbance caused by breath and heartbeat is another important factor that could lead to performance degradation. These disturbances could be partially corrected by applying the low pass filters. In the future, we will also extend our proposed weakly supervised framework to other OCT segmentation tasks, such as breast images and retinal images.

\section{Conclusion}
In this paper, we propose the first weakly supervised learning framework for adipose tissue segmentation in human cardiac OCT images. We design a novel CAM-superpixel segmentation approach which converts the sparse CAM results into pseudo pixel-wise labels for training. In addition, we also present and analyze the necessity and effectiveness of proposed steps and loss functions. Experimental results on the human cardiac dataset demonstrate that our model achieves comparable performance with models trained under full masks, showing the learning capability of our proposed model on image-level labels. 



\bibliography{mybibliography}{}

\begin{thebibliography}{10}

\bibitem{achanta2010slic}
Radhakrishna Achanta, Appu Shaji, Kevin Smith, Aurelien Lucchi, Pascal Fua, and
  Sabine S{\"u}sstrunk.
\newblock Slic superpixels.
\newblock Technical report, 2010.

\bibitem{alam2020av}
Minhaj Alam, David Le, Taeyoon Son, Jennifer~I Lim, and Xincheng Yao.
\newblock Av-net: deep learning for fully automated artery-vein classification
  in optical coherence tomography angiography.
\newblock {\em Biomedical Optics Express}, 11(9):5249--5257, 2020.

\bibitem{ambrosi2012quantification}
Christina~M Ambrosi, Vadim~V Fedorov, Igor~R Efimov, Richard~B Schuessler, and
  Andrew~M Rollins.
\newblock Quantification of fiber orientation in the canine atrial pacemaker
  complex using optical coherence tomography.
\newblock {\em Journal of Biomedical Optics}, 17(7):071309, 2012.

\bibitem{ambrosi2009virtual}
Christina~M Ambrosi, Nader Moazami, Andrew~M Rollins, and Igor~R Efimov.
\newblock Virtual histology of the human heart using optical coherence
  tomography.
\newblock {\em Journal of Biomedical Optics}, 14(5):054002, 2009.

\bibitem{bonou2021cardiac}
Maria Bonou, Sophie Mavrogeni, Chris~J Kapelios, George Markousis-Mavrogenis,
  Constantina Aggeli, Evangelos Cholongitas, Athanase~D Protogerou, and John
  Barbetseas.
\newblock Cardiac adiposity and arrhythmias: the role of imaging.
\newblock {\em Diagnostics}, 11(2):362, 2021.

\bibitem{el2021posterior}
Mohammed El~Mahdiui, Judit Simon, Jeff~M Smit, Jurrien~H Kuneman, Alexander~R
  van Rosendael, Ewout~W Steyerberg, Rob~J van~der Geest, Lili Sz{\'a}raz,
  Szilvia Herczeg, N{\'a}ndor Szegedi, et~al.
\newblock Posterior left atrial adipose tissue attenuation assessed by computed
  tomography and recurrence of atrial fibrillation after catheter ablation.
\newblock {\em Circulation: Arrhythmia and Electrophysiology}, 14(4):e009135,
  2021.

\bibitem{fang2017automatic}
Leyuan Fang, David Cunefare, Chong Wang, Robyn~H Guymer, Shutao Li, and Sina
  Farsiu.
\newblock Automatic segmentation of nine retinal layer boundaries in oct images
  of non-exudative amd patients using deep learning and graph search.
\newblock {\em Biomedical Optics Express}, 8(5):2732--2744, 2017.

\bibitem{fleming2010real}
Christine~P Fleming, Hui Wang, Kara~J Quan, and Andrew~M Rollins.
\newblock Real-time monitoring of cardiac radio-frequency ablation lesion
  formation using an optical coherence tomography forward-imaging catheter.
\newblock {\em Journal of Biomedical Optics}, 15(3):030516, 2010.

\bibitem{gan2019characterization}
Yu~Gan, Theresa~H Lye, Charles~C Marboe, and Christine~P Hendon.
\newblock Characterization of the human myocardium by optical coherence
  tomography.
\newblock {\em Journal of Biophotonics}, 12(12):e201900094, 2019.

\bibitem{gan2016automated}
Yu~Gan, David Tsay, Syed~B Amir, Charles~C Marboe, and Christine~P Hendon.
\newblock Automated classification of optical coherence tomography images of
  human atrial tissue.
\newblock {\em Journal of Biomedical Optics}, 21(10):101407, 2016.

\bibitem{glorot2010understanding}
Xavier Glorot and Yoshua Bengio.
\newblock Understanding the difficulty of training deep feedforward neural
  networks.
\newblock In {\em Proceedings of the Thirteenth International Conference on
  Artificial Intelligence and Statistics}, pages 249--256. JMLR Workshop and
  Conference Proceedings, 2010.

\bibitem{goergen2012optical}
Craig~J Goergen, Harsha Radhakrishnan, Sava Sakad{\v{z}}i{\'c}, Emiri~T
  Mandeville, Eng~H Lo, David~E Sosnovik, and Vivek~J Srinivasan.
\newblock Optical coherence tractography using intrinsic contrast.
\newblock {\em Optics Letters}, 37(18):3882--3884, 2012.

\bibitem{guo2019automatic}
Menglin Guo, Mei Zhao, Allen~MY Cheong, Houjiao Dai, Andrew~KC Lam, and Yongjin
  Zhou.
\newblock Automatic quantification of superficial foveal avascular zone in
  optical coherence tomography angiography implemented with deep learning.
\newblock {\em Visual Computing for Industry, Biomedicine, and Art}, 2(1):1--9,
  2019.

\bibitem{gupta2002imaging}
Meghna Gupta, Andrew~M Rollins, Joseph~A Izatt, and Igor~R Efimov.
\newblock Imaging of the atrioventricular node using optical coherence
  tomography.
\newblock {\em Journal of Cardiovascular Electrophysiology}, 13(1):95--95,
  2002.

\bibitem{huang2020segmentation}
Ziyi Huang, Yu~Gan, Theresa Lye, Darnel Theagene, Spandana Chintapalli, Simeran
  Virdi, Andrew Laine, Elsa Angelini, and Christine~P Hendon.
\newblock Segmentation and uncertainty measures of cardiac substrates within
  optical coherence tomography images via convolutional neural networks.
\newblock In {\em 2020 IEEE 17th International Symposium on Biomedical Imaging
  (ISBI)}, pages 1--4. IEEE, 2020.

\bibitem{huang2020heterogeneity}
Ziyi Huang, Yu~Gan, Theresa Lye, Haofeng Zhang, Andrew Laine, Elsa~D Angelini,
  and Christine Hendon.
\newblock Heterogeneity measurement of cardiac tissues leveraging uncertainty
  information from image segmentation.
\newblock In {\em International Conference on Medical Image Computing and
  Computer-Assisted Intervention}, pages 782--791. Springer, 2020.

\bibitem{islam2020deep}
Mohammad~Shafkat Islam, Jui-Kai Wang, Samuel~S Johnson, Matthew~J Thurtell,
  Randy~H Kardon, and Mona~K Garvin.
\newblock A deep-learning approach for automated oct en-face retinal vessel
  segmentation in cases of optic disc swelling using multiple en-face images as
  input.
\newblock {\em Translational Vision Science and Technology}, 9(2):17--17, 2020.

\bibitem{jiang2020weakly}
Zhe Jiang, Zhiyu Huang, Bin Qiu, Xiangxi Meng, Yunfei You, Xi~Liu, Mufeng Geng,
  Gangjun Liu, Chuanqing Zhou, Kun Yang, et~al.
\newblock Weakly supervised deep learning-based optical coherence tomography
  angiography.
\newblock {\em IEEE Transactions on Medical Imaging}, 40(2):688--698, 2020.

\bibitem{khurshid2018frequency}
Shaan Khurshid, Seung~Hoan Choi, Lu-Chen Weng, Elizabeth~Y Wang, Ludovic
  Trinquart, Emelia~J Benjamin, Patrick~T Ellinor, and Steven~A Lubitz.
\newblock Frequency of cardiac rhythm abnormalities in a half million adults.
\newblock {\em Circulation: Arrhythmia and Electrophysiology}, 11(7):e006273,
  2018.

\bibitem{kingma2014adam}
Diederik~P Kingma and Jimmy Ba.
\newblock Adam: A method for stochastic optimization.
\newblock {\em arXiv preprint arXiv:1412.6980}, 2014.

\bibitem{kolesnikov2016seed}
Alexander Kolesnikov and Christoph~H Lampert.
\newblock Seed, expand and constrain: Three principles for weakly-supervised
  image segmentation.
\newblock In {\em European Conference on Computer Vision}, pages 695--711.
  Springer, 2016.

\bibitem{lee2021automated}
Kyungmoo Lee, Alexis~K Warren, Michael~D Abr{\`a}moff, Andreas Wahle, S~Scott
  Whitmore, Ian~C Han, John~H Fingert, Todd~E Scheetz, Robert~F Mullins, Milan
  Sonka, et~al.
\newblock Automated segmentation of choroidal layers from 3-dimensional macular
  optical coherence tomography scans.
\newblock {\em Journal of Neuroscience Methods}, 360:109267, 2021.

\bibitem{lehtinen2018noise2noise}
Jaakko Lehtinen, Jacob Munkberg, Jon Hasselgren, Samuli Laine, Tero Karras,
  Miika Aittala, and Timo Aila.
\newblock Noise2noise: Learning image restoration without clean data.
\newblock {\em arXiv preprint arXiv:1803.04189}, 2018.

\bibitem{9600862}
Chao Li, Haibo Jia, Jinwei Tian, Chong He, Fang Lu, Kaiwen Li, Yubin Gong,
  Sining Hu, Bo~Yu, and Zhao Wang.
\newblock Comprehensive assessment of coronary calcification in intravascular
  oct using a spatial-temporal encoder-decoder network.
\newblock {\em IEEE Transactions on Medical Imaging}, 41(4):857--868, 2022.

\bibitem{lo2020microvasculature}
Julian Lo, Morgan Heisler, Vinicius Vanzan, Sonja Karst, Ivana~Zadro
  Matovinovi{\'c}, Sven Lon{\v{c}}ari{\'c}, Eduardo~V Navajas, Mirza~Faisal
  Beg, and Marinko~V {\v{S}}aruni{\'c}.
\newblock Microvasculature segmentation and intercapillary area quantification
  of the deep vascular complex using transfer learning.
\newblock {\em Translational Vision Science \& Technology}, 9(2):38--38, 2020.

\bibitem{lye2019imaging}
Theresa~H Lye, Charles~C Marboe, and Christine~P Hendon.
\newblock Imaging of subendocardial adipose tissue and fiber orientation
  distributions in the human left atrium using optical coherence tomography.
\newblock {\em Journal of Cardiovascular Electrophysiology}, 30(12):2950--2959,
  2019.

\bibitem{masood2019automatic}
Saleha Masood, Ruogu Fang, Ping Li, Huating Li, Bin Sheng, Akash Mathavan,
  Xiangning Wang, Po~Yang, Qiang Wu, Jing Qin, et~al.
\newblock Automatic choroid layer segmentation from optical coherence
  tomography images using deep learning.
\newblock {\em Scientific Reports}, 9(1):1--18, 2019.

\bibitem{meiniel2016sparsity}
William Meiniel, Yu~Gan, Christine~P Hendon, Jean-Christophe Olivo-Marin,
  Andrew Laine, and Elsa~D Angelini.
\newblock Sparsity-based simplification of spectral-domain optical coherence
  tomography images of cardiac samples.
\newblock In {\em 2016 IEEE 13th International Symposium on Biomedical Imaging
  (ISBI)}, pages 373--376. IEEE, 2016.

\bibitem{mirshahi2021foveal}
Reza Mirshahi, Pasha Anvari, Hamid Riazi-Esfahani, Mahsa Sardarinia, Masood
  Naseripour, and Khalil~Ghasemi Falavarjani.
\newblock Foveal avascular zone segmentation in optical coherence tomography
  angiography images using a deep learning approach.
\newblock {\em Scientific Reports}, 11(1):1--8, 2021.

\bibitem{pabon2018linking}
Maria~A Pabon, Kevin Manocha, Jim~W Cheung, and James~C Lo.
\newblock Linking arrhythmias and adipocytes: insights, mechanisms, and future
  directions.
\newblock {\em Frontiers in Physiology}, page 1752, 2018.

\bibitem{PEKALA2019103445}
M.~Pekala, N.~Joshi, T.Y.~Alvin Liu, N.M. Bressler, D.~Cabrera DeBuc, and
  P.~Burlina.
\newblock Deep learning based retinal oct segmentation.
\newblock {\em Computers in Biology and Medicine}, 114:103445, 2019.

\bibitem{pekala2019deep}
Mike Pekala, Neil Joshi, TY~Alvin Liu, Neil~M Bressler, D~Cabrera DeBuc, and
  Philippe Burlina.
\newblock Deep learning based retinal oct segmentation.
\newblock {\em Computers in Biology and Medicine}, 114:103445, 2019.

\bibitem{ronneberger2015u}
Olaf Ronneberger, Philipp Fischer, and Thomas Brox.
\newblock U-net: Convolutional networks for biomedical image segmentation.
\newblock In {\em International Conference on Medical Image Computing and
  Computer-Assisted Intervention}, pages 234--241. Springer, 2015.

\bibitem{samanta2016role}
Rahul Samanta, Jim Pouliopoulos, Aravinda Thiagalingam, and Pramesh Kovoor.
\newblock Role of adipose tissue in the pathogenesis of cardiac arrhythmias.
\newblock {\em Heart Rhythm}, 13(1):311--320, 2016.

\bibitem{sung2020personalized}
Eric Sung, Adityo Prakosa, Konstantinos~N Aronis, Shijie Zhou, Stefan~L
  Zimmerman, Harikrishna Tandri, Saman Nazarian, Ronald~D Berger, Jonathan
  Chrispin, and Natalia~A Trayanova.
\newblock Personalized digital-heart technology for ventricular tachycardia
  ablation targeting in hearts with infiltrating adiposity.
\newblock {\em Circulation: Arrhythmia and Electrophysiology}, 13(12):e008912,
  2020.

\bibitem{wang2011vivo}
Hui Wang, Wei Kang, Austin~P Bishop, Andrew~M Rollins, Thomas Carrigan, Noah
  Rosenthal, and Mauricio Arruda.
\newblock In vivo intracardiac optical coherence tomography imaging through
  percutaneous access: toward image-guided radio-frequency ablation.
\newblock {\em Journal of biomedical optics}, 16(11):110505, 2011.

\bibitem{wang2021weakly}
Jing Wang, Wanyue Li, Yiwei Chen, Wangyi Fang, Wen Kong, Yi~He, and Guohua Shi.
\newblock Weakly supervised anomaly segmentation in retinal oct images using an
  adversarial learning approach.
\newblock {\em Biomedical Optics Express}, 12(8):4713--4729, 2021.

\bibitem{yao2016myocardial}
Xinwen Yao, Yu~Gan, Charles~C Marboe, and Christine~P Hendon.
\newblock Myocardial imaging using ultrahigh-resolution spectral domain optical
  coherence tomography.
\newblock {\em Journal of Biomedical Optics}, 21(6):061006, 2016.

\bibitem{yoo2021feasibility}
Tae~Keun Yoo, Joon~Yul Choi, and Hong~Kyu Kim.
\newblock Feasibility study to improve deep learning in oct diagnosis of rare
  retinal diseases with few-shot classification.
\newblock {\em Medical \& Biological Engineering \& Computing}, 59(2):401--415,
  2021.

\bibitem{zhou2016learning}
Bolei Zhou, Aditya Khosla, Agata Lapedriza, Aude Oliva, and Antonio Torralba.
\newblock Learning deep features for discriminative localization.
\newblock In {\em Proceedings of the IEEE Conference on Computer Vision and
  Pattern Recognition}, pages 2921--2929, 2016.

\end{thebibliography}
\bibliographystyle{plain}

\end{document}